\documentclass[a4paper,twoside,twocolumn,english,showpacs,PRA]{revtex4}
\usepackage{times}
\usepackage[OT1]{fontenc}
\usepackage[latin1]{inputenc}
\usepackage{amsmath}
\usepackage{graphicx}
\usepackage{amssymb}

\makeatletter

\usepackage{babel}
\makeatother
\begin{document}

\title{Critical properties of a trapped interacting Bose gas}

\author{A. Bezett}
\author{P. B. Blakie}

\affiliation{Jack Dodd Centre for Quantum Technology,
Department of Physics, University of Otago, PO Box 56, Dunedin, New Zealand}
\date{\today}

\newcommand{\etal}{\emph{et al.}} 
\newcommand{\rC}{\text{\bf{C}}} 
\newcommand{\rI}{\text{\bf{I}}} 
\newcommand{\PC}{\mathcal{P}_{\rC}}
\newcommand{\bef}{\hat{\psi}}
\newcommand{\bfI}{\bef_{\rI}}
\newcommand{\cf}{\psi_{\rC}} 
\newcommand{\cfp}{\psi_{\rC'}} 
\newcommand{\ecut}{\epsilon_{\rm cut}} 
\newcommand{\CF}{c-field}
\newcommand{\Nc}{N_{\rm{cond}}}
\newcommand{\psic}{\psi_{\rm{cond}}}
\newcommand{\ac}{\alpha_{\rm{cond}}}
\newcommand{\thold}{t_{\rm{hold}}}
\newcommand{\nmin}{n_{\min}}
\def\x{\mathbf{r}} 
\newcommand{\xa}{(\x)} 

\begin{abstract}
We develop a practical theoretical formalism for studying the critical properties of a trapped Bose-Einstein condensate using the projected Gross-Pitaevskii equation. We show that this approach allows us investigate the behavior of the correlation length, condensate mode and its number fluctuations about the critical point. Motivated by recent experiments [Science {\bf 315}, 1556 (2007)] we calculate the critical exponent for the correlation length, observe clear finite-size effects, and develop characteristic length scales to characterize the finite-size influences.
We extend the Binder cumulant to the trapped system and discuss an experimental method for measuring number fluctuations.
\end{abstract}

\maketitle
\section{Introduction}
Phase transitions normally arise out of the competition between thermal fluctuations and inter-particle interactions. An important counterexample is that of the ideal Bose gas, where the second order transition to the condensed phase occurs in the absence of interactions, being driven solely by quantum statistical effects. However, the inclusion of even very weak interactions into the description of the condensation transition has proven to be a great theoretical challenge. For example, agreement on the effect of vanishingly small s-wave interactions on the critical temperature was only recently reached, after approximately 50 years of debate (e.g. see \cite{Andersen2004a}). 

Perhaps one of the most beautiful results of statistical physics is the concept of  universality -- that the properties of a system in the critical region are independent of the microscopic details of the system, and depend only on a few general system features such its dimensionality and the symmetry of the order parameter that emerges at the transition. In this context, the critical behavior of a weakly interacting 3D Bose gas should be identical to that of $^{4}$He at the superfluid transition, or to other systems of the same universality class (usually referred to as the 3D XY universality class). Accordingly, high precision experimental measurements of  critical $^{4}$He (e.g. see \cite{Lipa2003a}) are usually compared against theoretical calculations using idealized XY or $\phi^4$ models \cite{Kleinert1999a,Campostrini2001a,Burovski2006a}, rather than a microscopic description of the system.

The availability of experimental techniques for measuring correlations \cite{Schellekens2005a,Ottl2005a,Jeltes2006a,Folling2005a,Greiner2005a,Rom2006a} is an important feature of the ultra-cold atom systems that has received extensive theoretical attention, particularly in relation to (zero temperature) quantum phase transitions (e.g. see \cite{Altman2004a,Rey2006a,Rey2006b,Rey2006c,Ashhab2005,Scarola2006,Niu2006a,Hou2008,Toth2008a}). Another area of interest is the effect of critical fluctuations on correlations in the system at the finite temperature transition, particularly as other quantities usually examined in condensed matter (e.g. susceptibilities and heat capacity) are not easy to measure in atomic gases.
 While the critical fluctuations of weakly interacting 1D and 2D  Bose gases are dominant over a wide temperature range (e.g. see \cite{matterwave1,Polkovnikov}), it was previously thought that the width of the critical region about condensation temperature, $T_c$, would be far too narrow to permit experimental investigation in the 3D system.
However, in extraordinary recent experiments  Donner \etal\ \cite{critical} have made such measurements of a trapped 3D Bose gas, and were able to determine the critical exponent for the divergence of the correlation length to be $\nu=0.67\pm0.13$.  In its own right this result is an impressive demonstration that universality applies to a mesoscopic system with of order $10^5$ atoms. Additionally, this direct measurement of two-point correlations is of interest because it has not been possible to do this in helium, although a value for the correlation length critical exponent of helium ($\nu=0.67056\pm0.0006$) is inferred from the  heat capacity exponent using Josephson scaling relation, $\alpha=2-3\nu$.

\begin{figure}
\includegraphics[width=3.3in, keepaspectratio]{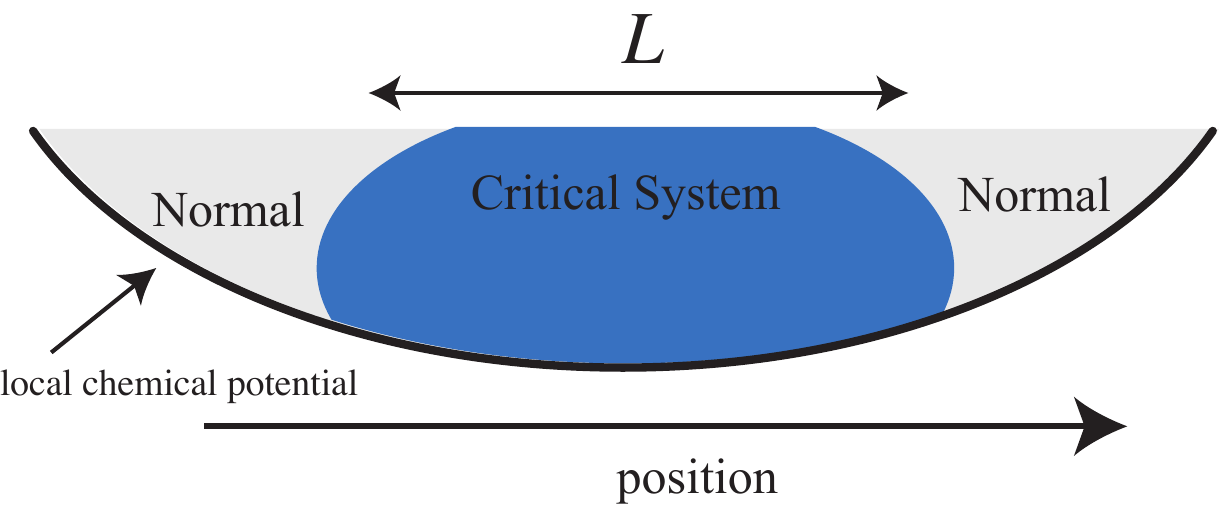}
\caption{\label{criticaltrap}  Schematic view of the critical regime of a trapped Bose gas. The local chemical potential varies across the system due to the harmonic trap potential. When the system is critical at the center, the criticality extends over the finite spatial range $L$ with a normal cloud boundary condition.}
\end{figure}

Finite-size effects have a profound influence on the critical properties of a system, and have been extensively studied to understand the cross-over of helium critical behavior during dimensional reduction \cite{Gasparini2008a}. Such systems, confined to a finite region of size $L$, are well-described by finite-size scaling theory \cite{Fisher1972a,Fisher1974a}. This theory shows that there is a universal scaling function that relates physical quantities of the finite to infinite systems depending on the quantity, the ratio of the correlation length to the system size, and the nature of the boundary conditions. 
In this context the scenario occurring in the harmonic trap is rather interesting (see Fig. \ref{criticaltrap}), and was first considered by Damle \etal\ in 1996 \cite{Damle1996a}. The effect of the trapping potential is to slowly vary the local value of the chemical potential. If the gas is critical at trap center, then moving out radially, the system gradually becomes normal. Thus the finite-size boundary conditions are rather difficult to describe, as they require an understanding of the (non-universal) normal system. The experiment by Donner \etal\ \cite{critical} did not observe finite-size effects:   their two-point measurements were made over a region much smaller than the spatial extent of the critical region, and  yielded a value of the critical exponent in line with the uniform system. 

In our opinion the study of the finite-size effects in this system should be rich and would be worth additional theoretical and experimental investigation.  However, a major impediment to the theoretical development is the lack of a practical formalism for studying the harmonically trapped Bose gas in the critical regime.  In this paper we report on progress towards such a formalism. We develop the projected Gross-Pitaevskii equation (PGPE) for application to the critical regime of the Bose-Einstein condensation transition. The PGPE method is a \emph{c-field} technique \cite{cfieldRev2008} applicable to the study of finite temperature degenerate Bose gases. It includes interactions between low energy modes of the gas non-perturbatively  and is applicable in the critical region, e.g. see \cite{DavisTemp,bezett,Simula2006a,Simula2008a}. Indeed, PGPE predictions for the shifts in critical temperature \cite{DavisTemp,Davis06} are in good agreement with other theoretical predictions in the uniform case \cite{Andersen2004a} and experimental measurements in the trapped system \cite{Gerbier04}. While this formalism has successfully predicted equilibrium properties for a degenerate Bose cloud, there has been little quantitative work at the critical point. While quantum Monte Carlo methods are now available for the trapped Bose gas \cite{Holzmann2007b}, we expect that the simpler   PGPE method to be far more efficient, easy to apply, and hence widely used. Indeed, the PGPE method is essentially an inhomogeneous $\phi^4$ theory. In this context the applicability of the PGPE to the critical regime is hardly surprising.
 However, unlike helium, where such models are thought to be inapplicable except in the critical regime, the PGPE theory also provides an accurate description of the  (non-critical) condensed and normal behavior for the dilute Bose gas.

\begin{figure}
\includegraphics[width=3.3in, keepaspectratio]{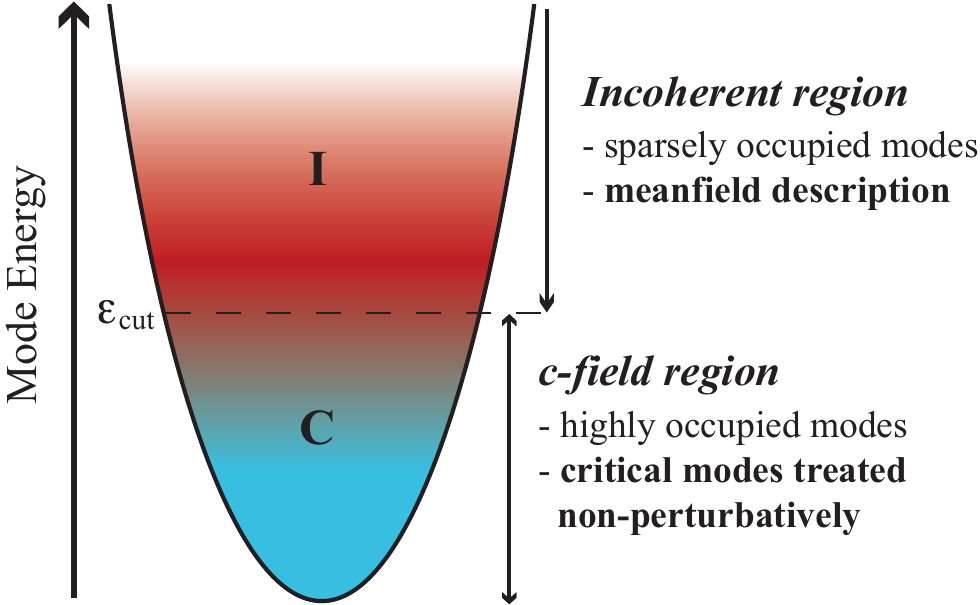}
\caption{\label{regions}  Schematic view of the \CF\ and the incoherent regions for a Bose gas in a harmonic trap potential, and the approximations we employ in our treatment of the collective mode dynamics.}
\end{figure}

The organisation of the paper is as follows. In Sec. \ref{Formalism} we review the projected Gross-Pitaevskii formalism, discussing how equilibrium states of the finite temperature Bose gas are generated and their properties are sampled. In Sec. \ref{Results} we present the main results of our research. We describe the detailed
macroscopic parameters of our simulations which span the critical point and present our results for the first order correlation functions and correlation length. In that section we also consider finite-size effects, and condensate number fluctuations. As far as we are aware, these are some of the first theoretical results systematically examining these quantities in the critical regime. Additionally, in the hope of stimulating further experimental work we present an observable, which we show is related to the condensate number fluctuations, and could be used to measure condensate statistics in experiment.
We conclude the paper in Sec. \ref{Conclusion}.
 
\section{Formalism}\label{Formalism}
We take our system to be described by  the second quantized Hamiltonian
\begin{eqnarray}
\hat{H} &=& \int  d\mathbf{r}\, \hat\Psi^{\dagger}(\mathbf{r})\left\{ H_0   + \frac{U_0}{2}\hat\Psi^{\dagger}(\mathbf{r})\hat\Psi(\mathbf{r})\right\}\hat\Psi(\mathbf{r}) ,
\end{eqnarray}
where 
\begin{equation}
H_0=-\frac{\hbar^2}{2m}\nabla^2+V_0(\mathbf{r}),
\end{equation}
is the single particle Hamiltonian, $\hat\Psi(\mathbf{r})$ is the quantum Bose field operator, and $U_0 = 4\pi\hbar^2a/m$ is the interaction strength, with $a$ the s-wave scattering length. The trap potential is given as
\begin{equation}
V_0(\mathbf{r}) = \frac{1}{2}m\left(\omega_x^2x^2+\omega_y^2y^2+\omega_z^2z^2\right).
\end{equation}

\subsection{PGPE formalism} \label{secPGPEformalism}
We briefly outline the projected Gross-Pitaevskii equation (PGPE) formalism, which is developed in detail in Ref. \cite{PGPE}.  
The Bose field operator is split into two parts according to
\begin{equation}
\hat\Psi(\mathbf{r}) = \cf(\mathbf{r}) + \bfI(\mathbf{r}),\label{EqfieldOp}
\end{equation}
where $\cf$ is the coherent region \CF\  and $\bfI$ is the incoherent field operator (see \cite{cfieldRev2008}).
These fields are defined as the low and high energy projections of the full quantum field operator, separated by the energy $\ecut$, as shown in Fig. \ref{regions}. In our theory this cutoff is implemented in terms of the harmonic oscillator eigenstates $\{\varphi_n(\mathbf{r})\}$ of the time-independent single particle Hamiltonian
i.e. $\epsilon_n\varphi_n(\mathbf{r})=H_0\varphi_n(\mathbf{r})$, with $\epsilon_n$ the respective eigenvalue.
The fields are thus defined by
\begin{eqnarray}
\cf(\mathbf{r}) &\equiv&\sum_{n\in\rC}c_n\varphi_n(\mathbf{r}),\label{Cfield}\\
\bfI(\mathbf{r}) &\equiv&\sum_{n\in\rI}\hat{a}_n\varphi_n(\mathbf{r}),
\end{eqnarray}
where the $\hat{a}_n$ are Bose annihilation operators, the $c_n$ are complex amplitudes, and the sets of quantum numbers defining the regions are 
\begin{eqnarray}
\rC &=&\{n:\epsilon_n\le \ecut\},\\
\rI &=&\{n:\epsilon_n> \ecut\}. 
\end{eqnarray} 
\subsubsection{Choice of $\rC$ region}
In general, the applicability of the PGPE approach to describing the finite temperature gas relies on an appropriate choice for $\ecut$, so that the modes at the cutoff have an average occupation of order unity. This choice means that the all the modes in $\rC$ are {appreciably occupied}, justifying the classical field replacement $\hat{a}_n\to c_n$. In contrast the $\rI$ region contains many sparsely occupied modes that are particle-like and would be poorly described using a classical field approximation. 
Because our interest here is in the critical regime, additional care is needed in choosing $\rC$. Typically strong fluctuations occur in the infra-red modes up to the energy scale  $U_0n$ where $n$ is the density. Above this energy scale the modes are well-described by mean-field theory (e.g. see the discussion in \cite{Kashurnikov2001a,Prokofev2001a}). 
For the results we present here, we choose to have 
\begin{equation}\ecut\sim k_BT> U_0n.\label{validitycond}
\end{equation}

\subsection{Equilibrium states}\label{secEqstates}
In this subsection we review our procedure for calculating finite temperature equilibrium properties of a trapped Bose gas. The basic approach is to treat the $\rC$ and $\rI$ regions as independent systems in thermal and diffusive equilibrium. We discuss the treatment of these regions separately below. Further details on this procedure are given in Sec. 3 of \cite{cfieldRev2008}.

\subsubsection{PGPE treatment of $\rC$ region}\label{SecformalismPGPE}

The equation of motion for $\cf$ is the PGPE
\begin{eqnarray}
i\hbar\frac{\partial \cf }{\partial t} = H_0\cf + \PC\left\{ U_0 |\cf|^2\cf\right\}, \label{PGPE}
\end{eqnarray}
where the projection operator 
\begin{equation}
\PC\{ F(\mathbf{r})\}\equiv\sum_{n\in\rC}\varphi_{n}(\mathbf{r})\int
d\mathbf{r}'\,\varphi_{n}^{*}(\mathbf{r}') F(\mathbf{r}'),\label{eq:projectorC}\\
\end{equation}
formalises our basis set restriction of $\cf$ to the $\rC$ region. The main approximation used to arrive at the PGPE is to neglect dynamical couplings to the incoherent region \cite{Davis2001b}.

An important feature of Eq.~(\ref{PGPE}) is that it is ergodic, so that  the microstates $\cf$ evolves through in time form a sample of the equilibrium microstates, and  time-averaging can be used to obtain macroscopic equilibrium properties.
Our basic procedure for finding equilibrium states consists of evolving the PGPE with three adjustable parameters: (i) the  cutoff energy, $\ecut$, that defines the division between $\rC$ and $\rI$, and hence the number of modes in the $\rC$ region; (ii) the number of $\rC$ region atoms, $N_{\rC}$; (iii) the total energy of the $\rC$ region, $E_\rC$. The last two quantities, defined as 
\begin{eqnarray}
E_{\rC}&=&\int d\mathbf{r}\,\cf^*\left(H_0 +  \frac{U_0}{2} |\cf|^2\right)\cf,\label{Ec}\\
N_{\rC} &=& \int d\mathbf{r}\,|\cf(\mathbf{r})|^2,
\end{eqnarray}
are important because they represent constants of motion of the PGPE (\ref{PGPE}), and thus control the equilibrium state of the system. 
\subsubsection{Randomized initial states}\label{randinitstates}
Given choices of these parameters, randomized initial states are constructed satisfying those constraints, and are evolved according to Eq.~(\ref{PGPE}). We now briefly describe how we produce these initial states.

We choose to make use of the Thomas-Fermi approximation to the condensate mode 
\begin{equation}
{\eta}_{\rm TF}(\mathbf{x})=\sqrt{\frac{\mu_{\rm TF}-{V}_{0}(\mathbf{x})}{U_0}}\,\,\theta\left(\mu_{\rm TF}-{V}_{0}(\mathbf{x})\right),
\end{equation}
where $\theta(x)$ is the unit step function and 
\begin{equation}
\mu_{\rm TF}=\frac{\hbar{\omega}}{2}\left(\frac{15 a N_{\rC}}{{a}_{\rm{ho}}}\right)^{2/5},\label{muTF}
\end{equation}   is the Thomas-Fermi chemical potential \cite{Dalfovo1999a}, with  ${\omega}=
(\omega_x\omega_y\omega_z)^{1/3}$  and ${a}_{\rm{ho}}=\sqrt{\hbar/m\omega}$.  We can generate a state of desired energy by superimposing the Thomas-Fermi state with a (high energy) randomized state, $\eta_{\rm{ran}}\xa$ (normalized to $N_{\rC}$), according to 
\begin{equation}
\eta_E\xa = p_0\eta_{\rm TF}\xa+p_1\eta_{\rm{ran}}\xa,\label{Estategen}
\end{equation}
where the variables $\alpha_0$ and $\alpha_1$ are adjusted to obtain the desired energy. In practice, $\eta_{\rm{ran}}$ is approximately orthogonal to $\eta_{\rm TF}$ and we can take $\alpha_1=\sqrt{1-|\alpha_0|^2}$. We note that since the randomized field only spans the $\rC$ region, it is most conveniently formed by the construction indicated in Eq. (\ref{Cfield}), but with the $c_n$ taken to be random complex numbers, globally scaled so that $\sum_{n\in\rC}|c_n|^2=N_{\rC}$.

 Typically evolution times of order 20 trap periods are used for the system to relax towards equilibrium \cite{Blakie2008a}, before properties of the equilibrium states are sampled using time-averaging, as we discuss below.
 
\subsubsection{Obtaining equilibrium properties for the $\rC$ region}\label{PGPEeqprops}
To characterize the equilibrium state in the $\rC$ region it is necessary to determine the average density, condensate fraction, temperature and chemical potential. Many of these quantities are also important for characterizing the $\rI$ region (see Sec. \ref{secMFrI}).  

The average density is obtained as a time average over the field microstates, i.e.,
\begin{eqnarray}
n_{\rC}(\mathbf{r})  &\approx& \frac{1}{M_s}\sum_{j=1}^{M_s}\left|\cf(\mathbf{r},t_j)\right|^2,\label{nc2}
\end{eqnarray}
where $\{t_j\}$ is a set of $M_s$ times (after the system has been allowed to relax to equilibrium) at which the field is sampled. We typically use $\sim7000$ samples over
the  140 trap periods  of our simulation to perform such averages.
The notation $n_\rC(\mathbf{r})$ emphasizes this is the density contribution from atoms residing in the $\rC$ region, with the obvious property that  $N_\rC=\int d\mathbf{r}\,n_\rC(\mathbf{r})$.

To find the mean condensate number, $\langle\Nc\rangle$ \footnote{We use the notation $\langle\Nc\rangle$ to indicate the mean condensate occupation to distinguish this from the equilibrium distribution of $\Nc$ we consider in Sec. \ref{Numflucts}.}, in our equilibrium state, we use the Penrose Onsager definition \cite{Penrose1956}, that  $\langle\Nc\rangle$ is given by the largest eigenvalue of the one-body density matrix 
\begin{equation}
G^{(1)}_{\rC}(\mathbf{r},\mathbf{r}') = \langle \cf^*(\mathbf{r})\cf(\mathbf{r}') \rangle,\label{G1}
\end{equation}
i.e.,
\begin{equation}
\int d\x' \,G^{(1)}_{\rC}(\mathbf{r},\mathbf{r}')\psic(\x')=\langle\Nc\rangle\psic\xa.
\end{equation}
Like the density, we can evaluate the one-body density matrix as a time-average.  Because the one-body density matrix  characterizes the emergence of the order parameter in the system at the critical point, we will examine $G^{(1)}_{\rC}(\mathbf{r},\mathbf{r}')$ further in Sec. \ref{spatcorrels} to characterize correlations in the system at the transition.

Derivatives of entropy, such as the temperature and chemical potential, are difficult to evaluate for interacting microcanonical systems. In 1997 Rugh proved that a microcanonical average (and hence time averaging) of appropriate quantities constructed from the Hamiltonian could be used to calculate entropy derivatives \cite{Rugh1997a}.  The Rugh method applies to classical mechanical systems, such as the PGPE which is the equation of motion for a system descrbed by the classical Hamiltonian (energy functional) (\ref{Ec}). The detailed implementation of the Rugh formalism for the PGPE is rather technical and we refer the reader to Refs. \cite{DavisTemp,Davis2005a} for additional details of this procedure.

\subsubsection{Meanfield treatment of the $\rI$ region}\label{secMFrI}
The average properties of the incoherent region can be calculated, at the level of Hatree-Fock meanfield theory, from
the one-particle Wigner distribution\begin{equation}
W_{\rI}(\mathbf{r},\mathbf{p})=\frac{1}{\hbar^3} \frac{1}{\exp(\beta[\epsilon_{\rm{HF}}(\mathbf{r},\mathbf{p})-\mu])-1},\label{eq:Wig3D}\end{equation} 
where 
\begin{eqnarray}
\epsilon_{\rm{HF}}(\mathbf{r},\mathbf{p}) & = & \frac{p^2}{2m}+ V_0(\mathbf{r})+2U_0(n_{\rC}(\mathbf{r})+n_{\rI}(\mathbf{r})),\label{eq:EHF3D}
\end{eqnarray}
is the Hartree-Fock energy, and $\mu$ is the chemical potential.  
In this semiclassical description $\mathbf{r}$ and $\mathbf{p}$ are treated as continuous variables. However, care needs to be taken to ensure that Eq.~(\ref{eq:Wig3D}) is only applied to the appropriate region of phase space spanned by the incoherent region,  i.e.~single-particle modes of energy exceeding $\ecut$. In phase space this region is  
\begin{equation}\label{WIdef}
\Omega_{\rI}=\left\{ \mathbf{r},\mathbf{p}:\frac{p^2}{2m}+V_0(\mathbf{r})\ge\ecut\right\} .
\end{equation}
This allows us to calculate density the incoherent region atoms
\begin{eqnarray}
n_{\rI}(\mathbf{r})&=& \int_{\Omega_\rI}{d\mathbf{p}}\, W_{\rI}(\mathbf{r},\mathbf{p}),
\end{eqnarray}
the number of incoherent region atoms $N_\rI  =\int d\mathbf{r}\,n_\rI(\mathbf{r})$, and thus the total number of atoms in the system $N=N_{\rC}+N_{\rI}$.

In Ref. \cite{bezett} it is shown the the full one-body density matrix for the system is given by
\begin{equation}
G^{(1)}(\x,\x') = G^{(1)}_{\rC}(\x,\x') + G^{(1)}_{\rI}(\x,\x'),\label{G1full}
\end{equation}
where $G^{(1)}_{\rC}(\x,\x')$  is given in Eq. (\ref{G1}) and
\begin{equation}
G^{(1)}_{\rI} (\x,\x') = \int_{\Omega_{\rI}} {d\mathbf{p}} \,e^{-i\mathbf{p}\cdot(\x -\x')/\hbar}\;W_{\rI}\left(\frac{\x+\x'}{2},\mathbf{p}\right).\label{G1wig}
\end{equation}
However, since we take $\ecut>U_0n$, then $G^{(1)}_{\rI}(\x,\x')$ contains only normal system correlations that decay on the length scale of the thermal de Broglie wavelength, $\lambda_{\rm{dB}}=h/\sqrt{2\pi mk_BT}$, so that the contribution of $G^{(1)}_{\rI}(\x,\x') $ is negligible for $|\x-\x'|\gtrsim \lambda_{\rm{dB}}$.

\section{Results}\label{Results}
\subsection{Sampling equilibrium states across the transition region}\label{EqStSampling} 
\begin{figure}
\includegraphics[width=3.3in, keepaspectratio]{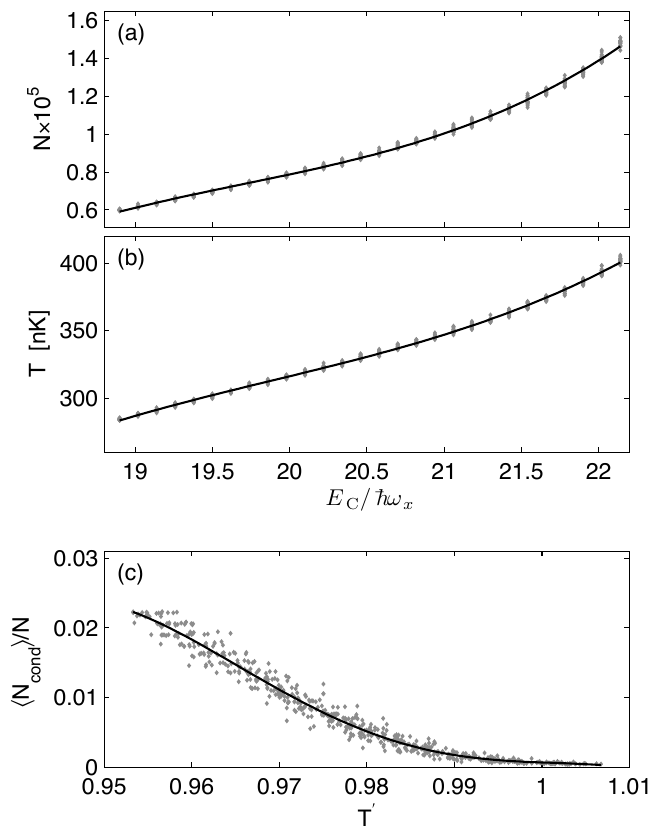}
\caption{\label{ParamsB} Macroscopic parameters for a critical trapped Bose gas.
(a) Total number of atoms and (b) temperature as the PGPE energy $E_{\rC}$ is changed. (c) Condensate fraction as a function of the reduced temperature $T/T_{c1}(N)$. Parameters: Rubidium-87 system with $\omega_{x,y}=2\pi\times129$ Hz, $\omega_z=2\pi\times364$ Hz, $\ecut=18.9\hbar\omega_x$ and $N_{\rC}=7573$. }
\end{figure}

We now discuss our procedure for generating equilibrium states spanning the condensation transition. We fix the variables $N_\rC$ and $\ecut$ to define our system and the generate equilibrium states with various energy values ($E_{\rC}$) finely spaced over a range where the thermalized condensate fraction is of order 1\%. 
Varying $E_{\rC}$ in this way causes $T$ to vary (as is desired), but also causes the total number of atoms to vary (see Fig. \ref{ParamsB}).

For each simulation we calculate the temperature and total atom number using the methods described in the previous section. The results for these quantities for the case
of a rubidium-87 system with $\omega_{x,y}=2\pi\times129$ Hz, $\omega_z=2\pi\times364$ Hz, $\ecut=32\hbar\omega_x$ and $N_{\rC}=7573$ are shown in Figs. \ref{ParamsB}(a) and (b). We have chosen to use these parameters rather than those of the ETH experiment \cite{critical} (which was carried out in a weaker trap), because the higher critical temperature of our parameters allows us to better satisfy the validity conditions of the PGPE theory (\ref{validitycond}).

For each energy we perform 20 simulations (using different random initial states) and the spread in results seen in Figs. \ref{ParamsB}(a) and (b) for each energy is indicative of the typical uncertainties in the thermal parameters. These results also show that as we change $E_{\rC}$ the total number of atoms in the system changes quite appreciably. It is therefore convenient to work in terms of the reduced temperature, $T'=T/T_{c1}$, where
\begin{eqnarray}
T_{c1} &=& T_{c0} - \left(0.73\frac{\bar{\omega}}{\omega}{N^{-\frac{1}{3}}} + 1.33\frac{a}{a_{\rm{ho}}}N^{\frac{1}{6}}\right)T_{c0}, \label{Tc1}
\end{eqnarray}
with 
\begin{equation}
{k_B}T_{c0} = 0.94\hbar\omega N^{1/3},
\end{equation}
  $\omega = (\omega_x \omega_y \omega_z)^{1/3}$, $\bar{\omega} = (\omega_x + \omega_y + \omega_z)/3$, and $a_{\rm{ho}} = \sqrt{\hbar/m\omega}$, see \cite{Giorgini1996a}.  The two terms in brackets in Eq. (\ref{Tc1}) correspond to the finite-size ($\propto N^{-1/3}$)  and meanfield interaction  ($\propto N^{1/6}$)  shifts of the critical temperature, respectively.
  
In Fig. \ref{ParamsB}(c) we show the condensate fraction from these simulations as a function of $T'$.  We note that $T'=1$ does not identify the transition precisely enough for understanding critical properties, as the above expression for $T_{c1}$ excludes meanfield effects beyond first order and does not account for any critical fluctuation effects.

\subsection{Spatial correlations and the correlation length}\label{spatcorrels}
In this section we consider the behavior of spatial correlations and the correlation length across the transition region.  
The spatial inhomogeneity of the trapped system requires explicit consideration. Unlike the homogeneous system, where spatial fluctuations are only a function of the separation between coordinates, in the trapped case both coordinates are separately important. So to study the development of order in the trapped system, we choose to examine correlations symmetrically about the trap center to minimize  inhomogeneous effects \footnote{In the experiments of Donner \etal\ correlations were also measured at points symmetrically placed about the trap center}. We do this by defining the normalized correlation function  
\begin{equation}
g_{\rC}(\Delta x) \equiv\frac{ G^{(1)}_{\rC}\left(\frac{1}{2}{\Delta x}{\,\hat{\mathbf{x}}},-\frac{1}{2}{\Delta x}{\,\hat{\mathbf{x}}}\right)}{n_{\rC}(\mathbf{0})},
\end{equation}
where $\hat{\mathbf{x}}$ is the unit vector in the $x$ direction, and $n_{\rC}(\mathbf{0})$ is the $\rC$ region density at trap center.
As discussed in Sec. \ref{secMFrI}, the spatial correlations over distances exceeding $\lambda_{\rm{dB}}$ are described completely by the $\rC$ region one-body density matrix.

\begin{figure}
\includegraphics[width=3.3in, keepaspectratio]{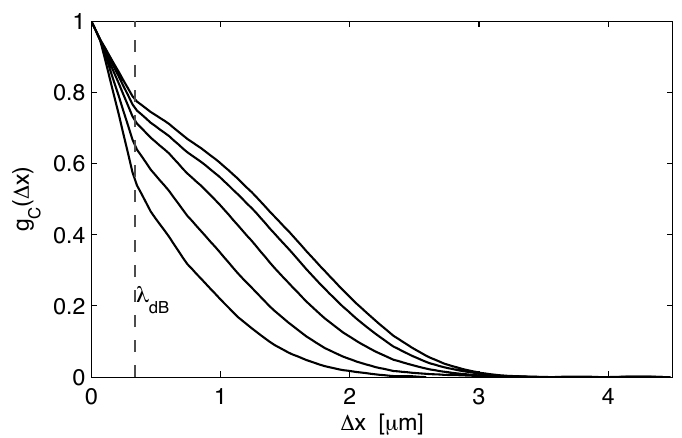}
\caption{\label{G1fig}  First order correlation function $g_{\rC}(\Delta x)$ for, from highest to lowest curves, $T'=\{0.957, 0.962, 0.969, 0.981, 0.990\}$. Parameters for calculations as in Fig. \ref{ParamsB}. Thermal de Broglie wavelength distance scale indicated for reference.}
\end{figure} 

 We evaluate $G^{(1)}_{\rC}(\x,\x')$ by time-averaging (see Sec. \ref{PGPEeqprops}), but due to our system's symmetry in the $xy$-plane, we can improve the quality of our results for $g_{\rC}(\Delta x)$ by making use of radial averaging.   Examples of $g_{\rC}(\Delta x)$ are shown in Fig. \ref{G1fig}.
In the region of the phase transition the first order correlation function is expected to take the form
\begin{equation}
g_{\rC}(\Delta x) \propto \frac{1}{\Delta x}e^{-\Delta x/\xi} \label{g1relate}
\end{equation}
for $\Delta x>\lambda_{\rm{dB}}$, where $\xi$ is the correlation length of the system. 
The variation in the correlation length as the temperature approaches the critical value is given by  
\begin{equation}
\xi \propto\left| {T'-T_c'}{}\right|^{-\nu}, \label{crit_exp}
\end{equation}
for the uniform system, 
where $\nu$ is the relevant critcal exponent.
 
\begin{figure}
\includegraphics[width=3.3in, keepaspectratio]{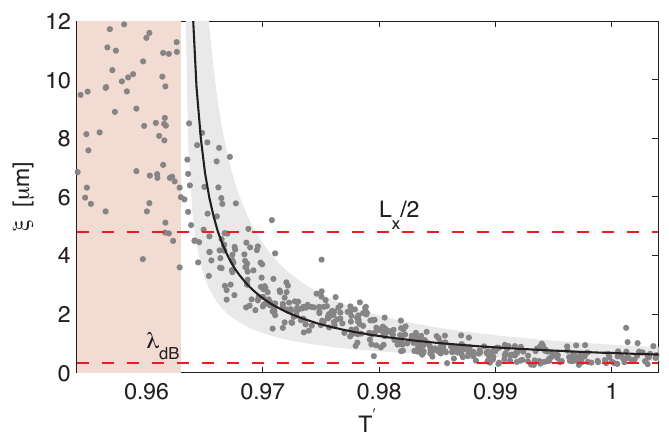}
\caption{\label{nufig}  The behavior of the correlation length across the condensation transition. Solid line is a fit using (\ref{crit_exp}) with $\nu=0.8$ and $T_c'=0.963$. Other parameters as in Fig. \ref{ParamsB}. Shaded region around the solid line indicates fits within the error bar range ($\nu=0.8\pm0.12$). Rectangular shaded region indicates points excluded as being below $T_c'$. Dashed lines show values of quantities discussed in the text.}
\end{figure}

To determine $\xi$  we fit (\ref{g1relate}) to our numerical results for $g_{\rC}(\Delta x)$ on the spatial range $0.5\mu$m $\le\Delta x\le2.2\mu$ m (i.e. $1.5\lambda_{\rm{dB}} \le\Delta x\le6.4\lambda_{\rm{dB}} $). This range matches that used by Donner \etal\ \cite{critical}, and ensures that we avoid having to deal with normal correlations at small separations, and inhomogeneous/finite-size effects at larger separations (also see in Sec. \ref{FSeffects}).
 The values of $\xi$ we obtain are shown in  Fig. \ref{nufig}, where we  see $\xi$ growing rapidly as $T'$ approaches $\sim0.96$ from above. At temperatures below this the quality of the fits used to determine $\xi$ is quite poor and there is appreciable scatter in the data points for $\xi$. This poor fit  arises from the development of appreciable condensate in the system  (e.g. see the coldest results shown in Fig. \ref{G1fig}). In the uniform system the condensate is spatially uniform and is easily neglected in correlation functions, however in the trapped system it appears at the transition point in a spatially localized mode with a size of order the oscillator length, which is difficult to distinguish from the non-condensate correlations.
We interpret the data for $T'\lesssim0.963$ as being below the transition point and our $\xi$ values extracted using fits to (\ref{g1relate}) in this regime as being unreliable.

We then fit expression (\ref{crit_exp}) to the correlation length results with $T_c'$, $\nu$, and an overall constant of proportionality as fitting parameters. The fit for our data is shown in Fig.  \ref{nufig}, with a value of $\nu=0.8\pm0.12$ and $T_c'=0.963$. We notice that while the fit is reasonable, there appears to be a certain degree of \emph{rounding off} in the divergence of $\xi$ near the critical point.
While our data has appreciable scatter (mainly due to uncertainty in $T$), we expect that this rounding off is primarily due to finite-size effects. Additionally, our large uncertainty in the critical exponent arises because of the difficulty in locating the precise value of $T_c$ for our system. In principle the divergence of $\xi$ marks $T_c$, however the  rounding off of this divergence and the scatter in our results adds uncertainties to the precise value of $T_c$ that is difficult to quantify without a theory for the finite-size behavior of $\xi$.

Our fit value for $\nu$ differs from that for the 3D XY model (i.e. $\nu\approx0.67$) which is expected to be of the same universality class,
 but is within the error bars of the Donner \etal\ experiment, which reported $\nu=0.67\pm0.13$.  
The critical temperature identified by our fit (i.e. $T'_c\approx 0.963$) is also shifted downward from the prediction of (\ref{g1relate}). A similar downward shift in $T_c$ was found by Davis \etal\ \cite{Davis06} in their analysis of the trapped Bose gas, arising because meanfield effects are typically underestimated by the analytic expression (\ref{Tc1}).

\subsubsection{finite-size effects}\label{FSeffects}
As our above results motivate, an important issue to consider in the trapped system is the role of finite-size effects \cite{Damle1996a}.
At fixed temperature, the Ginzburg criterion for the dominance of critical fluctuations requires that the chemical potential ($\mu$) differs from the critical value ($\mu_c$) by no more than
\begin{equation}
\delta\mu=|\mu-\mu_{c}|\le\frac{16\pi^{2}ma^{2}k_{B}^{2}T_{c0}^{2}}{\hbar^{2}},
\end{equation}
e.g. see \cite{Giorgini1996a}.  In the trapped system the effective system chemical potential varies spatially according to $\mu(\x)=\mu-V_0(\x)$ and, taking $\mu(\x=\mathbf{0})=\mu_c$, we can map the Ginzburg condition to a spatial length scale over which the system is critical. This length scale (diameter), along direction $x_j=\{x,y,z\}$, is
\begin{equation}
L_j=8\sqrt{2}\pi a\frac{k_{B}T_{c0}}{\hbar\omega_{j}},
\end{equation}
which we shall refer to as the Ginzburg length.
This sets the maximum correlation length that can occur in the system, thus defining the relevant parameter for assessing finite-size effects, and takes this (its largest) value when the center of the system is at the critical point. For the case of two-point correlations (measured along the $x$-direction for definiteness)  another important length scale in the trapped system is $\Delta x_{\max}$, the maximum point separation (about trap center) used to measure the correlation length (i.e. the fit of $\exp(-\Delta x/\xi)/\Delta x$ is made over the range $\lambda_{\rm{dB}}<\Delta x<\Delta x_{\max}$).

\begin{table}[htbp]
    \centering 
    \begin{tabular}{  c|ccc|c} 
     \hline
        System & $\quad\lambda_{\rm{dB}}(T_c)\quad$ & $\quad\Delta x_{\max}\quad$ & $\quad L_x\quad$ &$\nu$\\ 
       \hline     \hline
Expt. & $0.5\,\mu $m  & $2.2\,\mu $m & $20\,\mu $m & $0.67\pm0.13$  \\
Th.~1  & $0.34\,\mu $m  & $2.2\,\mu $m & $9\,\mu$m  &   $0.8\pm0.12$ \\
Th.~2  & $0.42\,\mu $m  & $2.2\,\mu $m & $6\,\mu $m  &  $0.8\pm0.12^\dagger$   \\
      \hline\hline
    \end{tabular}
    \vspace*{3mm}
    \caption{A summary of the parameters for critical property measurements. Expt. values refer to those of Donner \etal\ \cite{critical}. Th.~1 refer to the values for the main theoretical results presented in this paper and Th.~2 to the results presented in Sec. \ref{FSeffects}. $^\dagger$ These numbers are not fit, see text for additional discussion. }
    \label{tab:crit}
 \end{table}
 
The following conditions are required to accurately measure critical properties and minimise finite-size effects
\begin{equation}
\lambda_{\rm{dB}}\ll\Delta x_{\max}\ll L_j.\label{accreq}
\end{equation}
 The first inequality ensures that there is a reasonable distance scale over which correlation measurements can be made to accurately determine the correlation length. The second inequality ensures that finite-size effects are minimized. Obviously, finite-size effects cannot be completely avoided, since the correlation length can never diverge in the finite system, and $L_j$ sets the maximum value we might expect for $\xi$. The values of these various quantities for experiments and our results are shown in Table \ref{tab:crit}.

\begin{figure}
\includegraphics[width=3.3in, keepaspectratio]{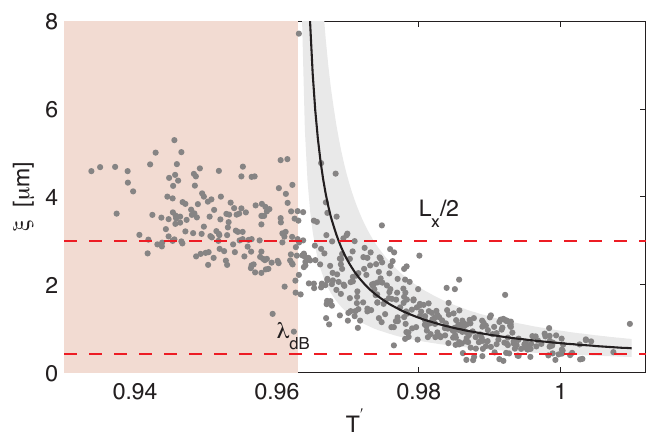}
\caption{\label{nufig2}The correlation length for a smaller system with $N_{\rC}=3.03\times10^3$ and $\ecut=23\hbar\omega_x$, with other parameters as  described in Sec. \ref{EqStSampling}. Solid line and surrounding shaded region are not fits  to the data, but are precisely the same as those used in Fig. \ref{nufig} (see text).
 Rectangular shaded region indicates points excluded as being below $T_c'$. Dashed lines show values of quantities discussed in the text.}
\end{figure}

 To examine the influence of finite-size effects, we have performed calculations for a system with the same trap parameters considered for the main results presented in this paper, but with fewer atoms. In this case the critical physics occurs at a temperature of  $T_c\approx200$ nK and a lower $\ecut$ value of  $23\hbar\omega_x$ is used. The relevant parameters for this system are summarized as ``Th.~2" in Table \ref{tab:crit}, revealing that for this system all three length scales in (\ref{accreq}) are similar. Our results for the correlation length behavior of this system are given in Fig.~\ref{nufig2} and show a rather striking broadening of the critical behavior, as compared to the previous case displayed in Fig. \ref{nufig}. As a result, this data is difficult to fit to the infinite system result (\ref{crit_exp}) for the purposes of extracting the critical exponent. Instead of fitting, we simply place the same curves used in Fig. \ref{nufig} (i.e. same $\nu$, $T_c'$, and error bars) on the data and observe that it provides an acceptable characterization of these results also.
  For both cases (Figs. \ref{nufig} and \ref{nufig2}) we see that fits to the normal divergent expression (\ref{crit_exp}) are good for $\xi\lesssim L_x/2$, but significantly depart from this fit for larger values of $\xi$. Since the values of $L_x$ differ by roughly a factor of two between these calculations,  this suggests that $L_x$ is indeed the correct length scale for assessing finite-size effects. 
 
\subsection{Condensate number fluctuations and the generalized Binder cumulant}\label{Numflucts}
An important issue to deal with in the trapped system is the identification of the critical point, as this will be needed for a better understanding of the critical region and the future development of higher precision calculations. For the uniform Bose gas the transition point is conveniently identified using a Binder cumulant, defined as
\begin{equation}
C_b=\frac{\langle N_0^2\rangle}{\langle N_0\rangle^2},
\end{equation}
where $N_0$ the population of the zero-momentum (condensate) mode. This Binder cumulant characterizes condensate  number fluctuations, and takes the universal value of $C_b^{\rm{crit}}=1.2430$ at the transition (see \cite{DavisTemp}).   

Here we propose a generalization of the Binder cumulant to the trapped system of the form
\begin{equation}
C_b^g \equiv \frac{\langle\Nc^2\rangle}{\langle\Nc\rangle^2},
\end{equation}
with $\Nc$ the condensate mode occupation. Our procedure to analyze the condensate number fluctuations is as follows. The condensate (lowest energy normal mode), $\psi_{\rm{cond}}\xa$,  is determined according to the Penrose-Onsager method described in Sec. \ref{SecformalismPGPE} using the time-averaged density matrix. We then use this mode to determine the instantaneous condensate amplitude by evaluating the inner product
\begin{equation}
\ac(t_j) = \int d\x\,\psic^*\xa\cf(\x,t_j),\label{alphac}
\end{equation}
on every microstate used to sample system properties.
We identify $\Nc=|\ac(t_j)|^2$ as the condensate number in this microstate, and by sampling over long times, we can obtain histograms of the condensate fluctuations.

\begin{figure}
\includegraphics[width=3.3in, keepaspectratio]{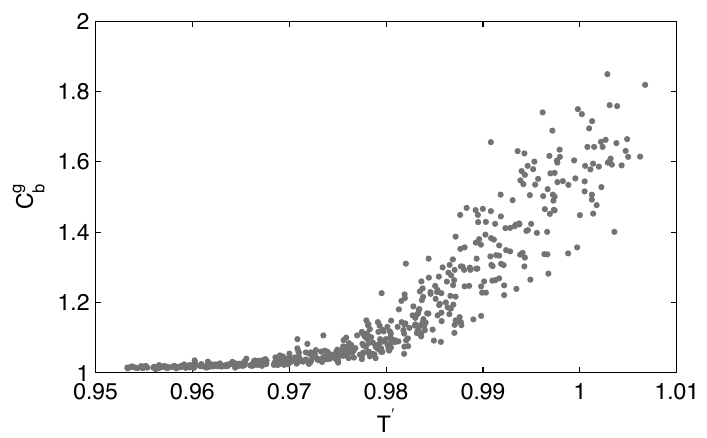}
\caption{\label{BinderCfig}  Binder cumulant behavior across the condensation transition. Parameters same as in Fig. \ref{ParamsB}}
\end{figure}
In Fig. \ref{BinderCfig} we show our results for $C_b^g$ across the critical region. These exhibit a rather dramatic reduction in the number fluctuations from values approaching $C_b^g=2$  (expected for the normal system) at temperatures slightly above the transition point, to values of $C_b^g\approx1$ below the transition.  At our transition temperature of $T_c'\approx0.963$ (as determined from fitting the critical exponent) we find a mean value of $C_b^g\approx1.03$, a value well-below the expected for the uniform system in the thermodynamic limit. This suggests that if the Binder cumulant is a useful quantity for characterizing the condensation transition in the trapped case, the critical value of $C_b^g$ is much lower than the uniform system.

\begin{figure}
\includegraphics[width=3.3in, keepaspectratio]{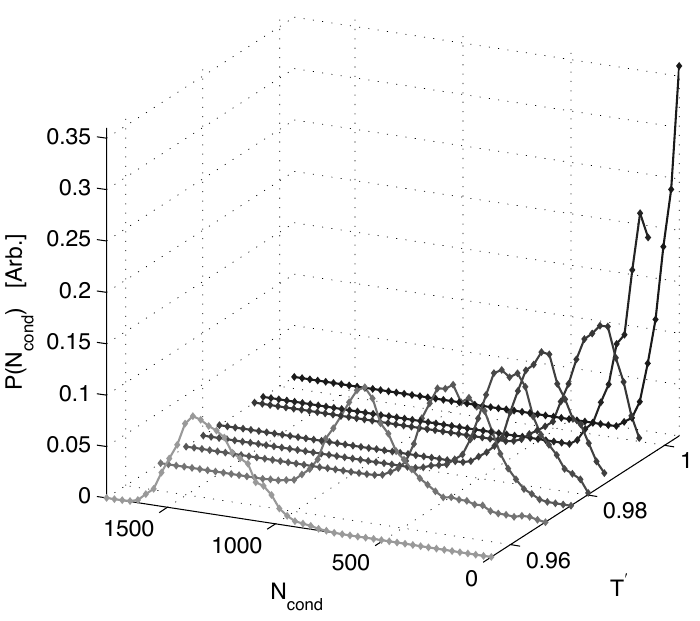}
\caption{\label{N0distfig}  Histograms of condensate/ground mode number fluctuations across the critical point.}
\end{figure}

Investigations of the number fluctuations of the condensate across the transition are of interest in their own right, and may be suitable to the techniques available in ultra-cold atom experiments. The subject of condensate number fluctuations has been extensively discussed in the dilute gas BEC literature (e.g. see the review of \cite{Kocharovsky2006a} and references therein), particularly the unphysical large fluctuations for the ideal gas predicted within the grand canonical ensemble. 

In Fig. \ref{N0distfig} we show histograms of the the condensate (lowest mode) number distribution across the transition. We see the development of coherence in the system as the shape of the distribution changes from being maximum at $\Nc=0$ above the transition, to having a maximum at finite $\Nc$ below the transition. We also find that as the temperature decreases the condensate number fluctuations are suppressed, i.e. 
\begin{equation}
\frac{\langle (\Nc-\langle \Nc\rangle)^2\rangle}{\langle \Nc\rangle^2} \to 0,
\end{equation}
(as was implicit in the behavior of $C_b^g$ observed above) and that the distribution has negative skew. 

Given the phenomenal recent interest in measuring spatial correlations in ultra-cold atom systems \cite{Schellekens2005a,Ottl2005a,Jeltes2006a,Folling2005a,Greiner2005a,Rom2006a}, it would be of great interest to develop analogous techniques for observing these condensate number distribution in experiments. 
It is difficult to devise an experimental procedure which could be used to measure $\ac$ (or $\Nc$) in a  manner equivalent to (\ref{alphac}), which requires complete phase and amplitude information about the field. So here we propose a quantity that can be directly measured in experiments and used to reveal the transition from incoherent to coherent number statistics of the condensate mode. In particular, we consider the central momentum column density
 \begin{equation}
n_{p=0}\equiv\left[\int dp_z\,n(\mathbf{p})\right]_{p_x=p_y=0},
\end{equation}
as an observable, since it proportional to the peak density measured in the usual absorption images taken of ultra-cold systems \cite{Ketterle1999a}. The motivation for choosing this quantity is that  the long range coherence of the condensate is clearly revealed as a peak in momentum space, thus the central momentum value is correlated with the condensate occupation. The detailed relationship between $n_{p=0}$ and $\Nc$ is not unique, due to the contribution of the noncondensate to $n_{p=0}$. So measurements of $n_{p=0}$ cannot be considered equivalent to the condensate yet, as we show below, there are clear qualitative similarities between the distributions of both quantities.

\begin{figure}
\includegraphics[width=3.3in, keepaspectratio]{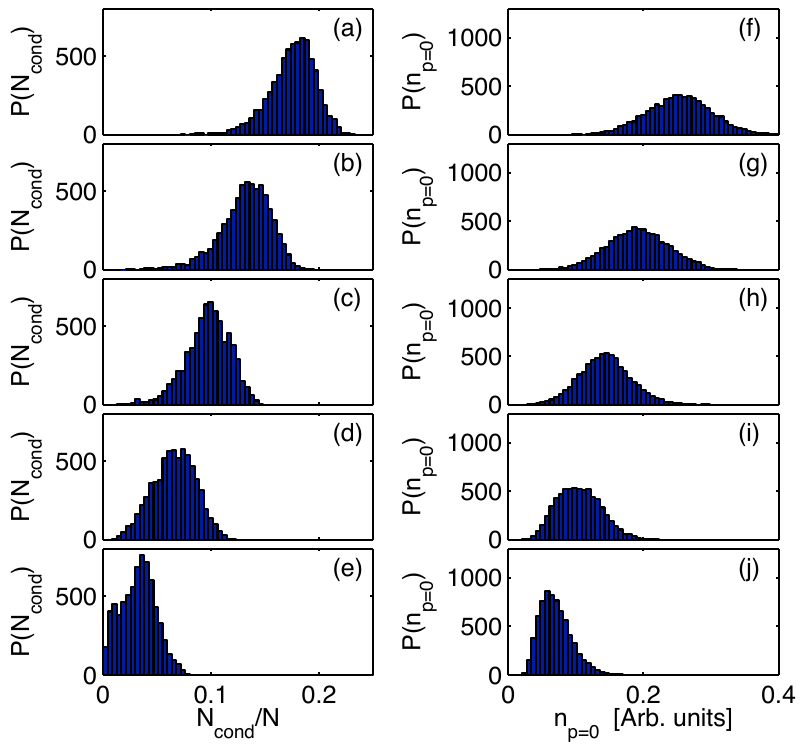}
\caption{\label{numfluctsfig}  Comparison of (a-e) histograms of condensate/ground mode number distribution and (f-j) the the central momentum column density distribution across the critical point. Results (a-e) and (f-j) correspond to the temperatures $T'=\{0.957, 0.962, 0.969, 0.981, 0.990\}$ respectively.}
\end{figure}

In Fig. \ref{numfluctsfig} we compare the distributions for $n_{p=0}$ and $\Nc$ obtained from analysis of data sets from the same PGPE calculations. 
Qualitatively, the behavior of these distributions appears to be quite similar. The $n_{p=0}$ distribution is clearly seen to be offset from zero at high temperatures as compared to the $\Nc$ distribution (see Figs. \ref{numfluctsfig}(e) and (j)). This offset is related to the average momentum column density of the noncondensate component.

\section{Conclusions and outlook}\label{Conclusion}
In this work we have developed a theoretical tool for studying critical physics in the 3D trapped Bose gas, motivated by the recent theoretical work by the ETH group \cite{critical}. For numerical convenience we have studied a trapped Bose gas with different parameters to the experimental system, yet obtain a critical exponent that agrees to within the error bars of both results. We have discussed finite-size effects and show that they can significantly alter the critical physics for systems with a small Ginzburg length.
Finally, we considered fluctuations of the condensate mode occupation across the transition region, and have shown that measurements of the central momentum column density can be used to experimentally reveal the emergence of coherent statistics in the system.

 Our study here represents the first quantitative theoretical calculations for this system beyond meanfield level. However, there are numerous avenues for improvement of our approach that could be used to  obtain higher precision results, which we believe will be needed to develop a better understanding of the trapped system critical physics, and stimulate more experiments in this fascinating new area for ultra-cold atomic gases.
  For instance, it would be interesting to apply the stochastic PGPE (SPGPE) formalism \cite{Gardiner2003a} to the critical regime, as both the chemical potential and temperature are control parameters in this approach. This may allow more precise measurement of critical exponents and would make direct comparisons with the fixed $N$ experiment of Donner \etal\ \cite{critical} more easy to achieve \footnote{i.e. it is quite difficult to analyze a system of fixed number and varying temperatures using the PGPE approach as both $N$ and $T$ vary as we change $E_{\rC}$.}. Another avenue of investigation would be to build on our formalism a more efficient method of sampling,  e.g. using Monte Carlo algorithms  (e.g. see \cite{Kashurnikov2001a}). With additional improvements in precision  we believe our formalism will be able to provide a detailed characterization of the finite-size cross over functions for the trapped Bose gas.  Knowledge of these functions would be useful for several problems of current interest in the ultra-cold atomic physics, such as better understanding of the quasi-2D behavior  and the emergence of phase defects in quenches across the critical regime \cite{Weiler08a}.

\begin{acknowledgments} 
AB acknowledges support of a TEC Top Achiever Doctoral Grant.
PBB wishes to acknowledge useful discussions with  M.~J.~Davis, and use of the Otago Vulcan Cluster.
This work was supported by the New Zealand Foundation for Research, Science and
Technology under Contract Nos.  NERF-UOOX0703.  \end{acknowledgments}

\end{document}